\documentclass[reqno]{amsart}
\usepackage{amsfonts}
\usepackage{amsthm}
\usepackage{amssymb}
\usepackage{dsfont}
\usepackage{amscd}
\usepackage{color}
\usepackage{tikz}
\usepackage{float}

 \usetikzlibrary{calc,decorations.markings,matrix,decorations.pathmorphing}

\newtheorem{theorem}{Theorem}[section]
\newtheorem{cor}[theorem]{Corollary}

\newtheorem{prop}[theorem]{Proposition}
\theoremstyle{definition}

\theoremstyle{remark}

\numberwithin{equation}{section}

\def\be{\begin{equation}}
\def\ee{\end{equation}}
\def\ba{\begin{eqnarray*}}
\def\ea{\end{eqnarray*}}
\def\bae{\begin{eqnarray}}
\def\eae{\end{eqnarray}}
\def\bc{\begin{center}}
\def\ec{\end{center}}

\begin{document}

\title[Finite size corrections and Odlyzko's data set for the Riemann zeros]{Finite size corrections in random matrix theory and Odlyzko's data set for the Riemann zeros}
\author{Peter J. Forrester and Anthony Mays} \address{Department of Mathematics and Statistics \\
ARC Centre of Excellence for Mathematical \& Statistical Frontiers\\
The University of Melbourne, Victoria 3010, Australia; }

\begin{abstract}
Odlyzko has computed a data set listing more than $10^9$ successive Riemann zeros, starting at a zero number beyond $10^{23}$. The data set relates to random matrix theory since, according to the Montgomery-Odlyzko law, the statistical properties of the large Riemann zeros agree with the statistical properties of the eigenvalues of large random Hermitian matrices. Moreover, Keating and Snaith, and then Bogomolny and collaborators, have used $N \times N$ random unitary matrices to analyse deviations from this law. We contribute to this line of study in two ways. First, we point out that a natural process to apply to the data set is to thin it by deleting each member independently with some specified probability, and we proceed to compute empirical two-point correlation functions and nearest neighbour spacings in this setting. Second, we show how to characterise the order $1/N^2$ correction term to the spacing distribution for random unitary matrices in terms of a second order differential equation with coefficients that are Painlev\'e transcendents, and where the thinning parameter appears only in the boundary condition. This equation can be solved numerically using a power series method. Comparison with the Riemann zero data shows accurate agreement.
\end{abstract}

\maketitle

\section{Introduction}
The celebrated Riemann hypothesis asserts that all the non-trivial zeros of the Riemann zeta function $\zeta(s)$
are on the so-called
critical line, Re$\, s = 1/2$. We will refer to zeros of $\zeta(s)$ on the critical line as Riemann zeros, and restrict attention
to those in the upper half plane --- those in the lower half plane are obtained by complex conjugation.
There is an intriguing relationship between the Riemann zeros
and random matrix theory.
It originates from the viewpoint that the  Riemann
zeros, for distances far up the critical line, are best considered in a statistical sense rather than as a deterministic sequence. 
The first step in a statistical analysis is to scale the sequence so that locally the density of the zeros far up the critical line is
unity. This is straightforward from knowledge of the fact that at position ${1 \over 2} + i E$ along the critical line,
and with $E \gg 1$, the density ($\bar{\rho}$ say) is given by \cite[pg.~280]{WW65}
\begin{equation}\label{1A}
\bar{\rho} = \frac1{2\pi} \log \Big ( {E \over 2 \pi e} \Big ) +  {\rm O} \Big ( {\log E \over E} \Big ).
\end{equation}
Montgomery \cite{Mo73,MS99} made a study of the density of the scaled zeros about a fixed zero, with the latter averaged over
some window of size $\Delta E$, with $\Delta E \ll E$. This quantity, referred to as the  \textit{two-point correlation function},
is probed by computing the average value of a test function $f(E-E_0)$, with $E_0$ corresponding to the fixed zero.
Subject to the technical assumption that the class of test function used has Fourier transform supported on $|k| < 2 \pi$,
the limiting two-point correlation function was proved to be equal to
\begin{equation}\label{S}
\lim_{N\to \infty} \rho^{\mathrm{R}}_{(2)} (E, E+ s) = 1 -  \Big ( {\sin \pi s \over \pi s} \Big )^2,
\end{equation}
where the superscript `R' denotes that this quantity pertains to the Riemann zeros data. On the other hand, let $X$ be an $N \times N$ random matrix with standard complex Gaussian entries, and form the Hermitian matrix ${1 \over 2} (X + X^\dagger)$ --- the set of such matrices specifies the Gaussian Unitary Ensemble (GUE). After normalisation so that the eigenvalues in the bulk of the spectrum have unit density, the
$N \to \infty$ form of the two-point correlation function is precisely (\ref{S}); see e.g.~\cite[Prop.~7.1.1]{Fo10}.

This coincidence, first observed by Dyson (see e.g.~\cite[pg.~159]{Bo05}), is in keeping with and extends the so-called \textit{Hilbert--P\'olya conjecture} \cite{WikiHP} that the Riemann zeros correspond to the eigenvalues of some unbounded self-adjoint operator. Considerations  in quantum chaos (see e.g.~\cite{Ha90,Bo00}) give evidence that generically 
the large eigenvalues of such operators have local statistics coinciding with the eigenvalues of large random Hermitian matrices. The latter must be constrained to have
real entries should there be a time reversal symmetry. Thus this line of reasoning suggests that asymptotically the
Riemann zeros have the same local statistical properties as the large eigenvalues of a chaotic quantum system without time reversal symmetry. This assertion is essentially a statement of what now is called the Montgomery--Odlyzko law
or the GUE conjecture; see also
\cite{Be86,BK99} the reviews \cite{Bo05,FM09a}, and the thesis \cite{Ro13}.

At an analytic level, further evidence for the validity of the law was given by Rudnick and Sarnak \cite{RS95}, who extended
Montgomery's result to the general $k$-point correlation function by obtaining the explicit functional form
\begin{equation}\label{S1}
\det \left[ {\sin \pi (x_j - x_l) \over \pi (x_j - x_l)} \right]_{j,l=1,\dots,k},
\end{equation}
again subject to a constraint on the Fourier transform of the class of test functions involved;
see the recent work \cite{CS14} for some refinements. By use of a conjecture of Hardy and Littlewood relating to the pair correlation of prime numbers, and also at the expense of introducing some non-rigorous working, Bogomolny and Keating \cite{BK95, BK96a, BK96b} removed this constraint. The $k$-point correlation (\ref{S1}) is precisely that for the eigenvalues of large Hermitian random matrices in the bulk scaling limit; see again~\cite[Prop.~7.1.1]{Fo10}.
At a numerical level, Odlyzko has made high precision computations of the $10^{20}$-th Riemann zero and over 70 million of its neighbours \cite{Od87}, and (later) of the $10^{22}$-nd zero and one billion of its neighbours \cite{Od01}. (The zeros on the critical line are conventionally numbered in the order of their distance from the real axis, with the first zero at $\mathrm{Im}\, s\approx 14.134725$ \cite[A058303]{OEIS}.) This data exhibits consistency with random matrix theory for statistical quantities such as the pair correlation function outside the range known rigorously from the work of Montgomery,  the distribution of the spacing between neighbouring zeros, the variance of the fluctuation of the number density in an interval etc.; for a popular account of this line of research see \cite{Ha03}.

Notwithstanding the extraordinary distance along the critical line achieved in Odlyzko's calculations, it turns out that finite size corrections to the limiting behaviour occur on the scale of the logarithm of the distance as suggested by
(\ref{1A}), and thus are of significance in the interpretation of the data from the viewpoint of the random matrix predictions. In addressing the question of the functional form of the finite size corrections from the viewpoint of random matrix theory, Keating and Snaith \cite{KS00a} (see also the review \cite{KS03} and the later work \cite{CFKRS08}) were led to a totally unexpected conclusion: the correct model for this purpose is not complex random Hermitian matrices, but rather $N \times N$ random unitary matrices chosen with Haar measure,
and $N$ related to $E$ so as to be consistent with (\ref{1A}).
Random unitary matrices with Haar measure result, for example, by applying the Gram-Schmidt orthonormalisation procedure to a matrix of standard complex Gaussians. The statistical quantities considered in \cite{KS00a} relate to the value distribution of the logarithm of the Riemann zeta function on the critical line, which for random matrices corresponds to the value distribution of the characteristic polynomial. 
Coram and Diaconis \cite{CD03} subsequently used the random unitary matrix model to predict the functional form
of the covariance between the number of eigenvalues in overlapping intervals of equal size.
Following up on  \cite{Od01}, where Odlyko studies the difference between the spacing distribution for the Riemann zeros
and the corresponding limiting random matrix distribution and
 writes: ``Clearly there is structure in this difference graph, and the challenge is to understand where
it comes from", 
Bogomolny et al.~\cite{BBLM06} considered finite size corrections to the spacing distribution 
for random unitary matrices.
While previous studies in random matrix theory gave analytic results relating to the value distribution of the characteristic polynomial \cite{BF97f}, there is no existing literature on the analytic calculations of the leading finite size correction to the spacing distribution for random unitary matrices. 
Here we are referring to the point-wise value of the expected probability density function; there is an existing literature relating
to the decay as a function of $N$ of the Kolmogorov distance between the empirical cumulative spacing distribution and
its expected large $N$ form \cite{KS99,Sc12,KS13}.
In \cite{BBLM06} the corrections were computed using an extrapolation procedure of a determinant expression valid for finite $N$. A primary objective of the present paper is to provide an analytic characterisation of the leading finite size correction to the spacing distribution.

Variants of the nearest neighbour spacing distribution also provide relevant statistical quantities. For example, one could consider
spacing distributions 
between zeros $k$ apart (see e.g.~\cite[\S 8.1]{Fo10}),
or the distribution of the closest of the left and right neighbours \cite{FO96}.
Another variant is to consider spacing distributions resulting from first thinning the
data set by the process of deleting each member independently with probability $(1 - \xi)$, $0 < \xi < 1$,
as first considered in random matrix theory by Bohigas and Pato \cite{BP04}. In both cases the
finite size correction to the large $N$ form of the corresponding quantity for the eigenvalues of random unitary matrices can be
computed analytically.  This is of concrete consequence as we are fortunate enough to have had  A.~Odlyzko provide us with a
data set extending that reported in \cite{Od01}. Specifically, this data set begins with zero number
$$
100,000,000,000,000,985,531,550 \approx 10^{23},
$$ 
and this occurs at the point $s = 1/2 + iE$ in the complex $s$-plane with $E$ equal to
\begin{align}
\nonumber 13066434408793621120027.3961465854 \approx 1.30664344 \times 10^{22}.
\end{align}
The data set contains a list of slightly more than $10^9$ of each of the subsequent Riemann zeros. Thus we are able to compare the analytic forms against the finite corrections exhibited by the Riemann zeros, uninhibited by sampling error.

We begin in Section \ref{S2} by considering the two-point correlation. Bogomolny and Keating \cite{BK96b} have derived an
explicit formula for this quantity in the case of the Riemann zeros, in the regime that $E$ is large but finite.
Subsequently Bogomolny et al.~\cite{BBLM06} showed how the result of \cite{BK96b} could be expanded for large $E$
to obtain the correction term to the random matrix result (\ref{1A}). Moreover, upon (partial) resummation,
the correction term was found to be identical in its functional form to the correction term of the scaled large $N$ expansion
of the two-point correlation for the eigenvalues of random unitary matrices chosen with Haar measure. After reviewing
this result, we compare the empirical form of the two-point correlation as computed from Odlykzko's data set, and for the
data set thinned with parameter $\xi = 0.6$, against the theoretical prediction. Very good agreement is found for
distances up to approximately one and a half times the average spacing between zeros, after which systematic deviation is
observed.

Higher order correlations for random unitary matrices are fully determined by the same function --- see (\ref{2.0}) below --- as determines the two-point correlation.
Assuming the same is true for the correlations of the Riemann zeros, it follows that the coincidence between the
correction term to the leading large $E$ form of the two-point correlation for the Riemann zeros and the
correction term to the leading large $N$ form of the two-point correlation for the eigenvalues of random unitary matrices
persists to all higher order correlations. While higher order correlations cannot be measured
empirically from the Riemann zeros data, this hypothesis becomes predictive as the coincidence must
carry over to any distribution function which can be written in terms of the correlation functions, for example the
nearest neighbour spacing distribution  \cite{BBLM06}. In Section 3 we make two main contributions to the study of this theme.
One is to test the hypothesis upon thinning of the data set. The other is to provide an analytic determination of the
correction term in the random matrix case, using certain Painlev\'e transcendents. 

The correction term to the nearest neighbour spacing is oscillatory, and becomes more so as the thinning parameter
$\xi$ is decreased from 1. In Sections \ref{s:space} and \ref{s:Num} we discuss this effect in the context of the large distance form of the Painlev\'e transcendents.

\section{Two-point correlation}\label{S2}

Correlation functions are fundamental to the theoretical description of general point processes. For definiteness, and in keeping with the setting of the Riemann zeros, we will specify that the point process is defined on a line. Starting with $\rho_{(1)}(x)$ as the density at point $x$, the $k$-point correlations $\rho_{(k)} (x_1, \dots,x_k)$ can be defined inductively by the requirement that the ratio
$$
{\rho_{(k)}(x_1,x_2,\dots,x_k) \over \rho_{(k-1)}(x_1,x_2,\dots,x_{k-1})}
$$
corresponds to the density at the point $x_k$, given that there are particles (or zeros, or eigenvalues etc.) at locations $x_1,\dots,x_{k-1}$. Suppose the particle density $\rho_{(1)} (x)$ is identically constant so that the point process is translationally invariant, and furthermore take the constant to be unity. We then have that the two-point correlation $\rho_{(2)} (x_1, x_2)$ is just the density at $x_2$ given that there is a particle at $x_1$. As such, in this circumstance, $\rho_{(1)} (x_1, x_2)$ can be empirically determined from a single data set. To see this note that due to translational invariance, $\rho_{(2)}(x_1,x_1+s)$ depends only on $s$, allowing $x_1$ to be averaged over to generate the necessary data for a statistical determination.

For the eigenvalues $\{ e^{i \theta_j} \}_{j=1,\dots,N}$ of $N \times N$ unitary matrices chosen with Haar measure, the $k$-point correlation is given by the $k \times k$ determinant (see e.g.~\cite[\S 5.5.2]{Fo10})
\begin{align}
\nonumber \rho_{(k)} (x_1, \dots, x_k) = \det [ K_N(x_i,x_j) ]_{i,j=1,\dots,k}
\end{align}
where $x_j = N \theta_j/ 2 \pi$ ($j=1,\dots,N$) and $K_N(x,y)$ --- referred to as the \textit{correlation kernel} --- is given by
\begin{equation}\label{2.0}
K_N(x,y) = {\sin \pi (x - y) \over N \sin (\pi (x-y)/N)}.
\end{equation}
In particular the two-point correlation is given by
$$
\rho_{(2)}(x,x+s) = 1 - \left( {\sin \pi s \over N \sin \pi s/N} \right)^2,
$$
which one sees tends to (\ref{S}) in the limit $N \to \infty$. We are particularly interested in the leading correction term to this asymptotic form. As reported in \cite{BBLM06}, an elementary calculation gives
\begin{equation}\label{Sc}
\rho_{(2)}(x,x+s) = 1 - \Big ( {\sin \pi s \over \pi s} \Big )^2 - {1 \over 3 N^2} \sin^2 \pi s +
{\rm O} \Big ( {1 \over N^4} \Big ).
\end{equation}

The situation with the Riemann zeros is more complicated. For a start, since the Riemann zeros are a deterministic sequence, a statistical characterisation only results after an averaging over a suitable interval of zeros about the point $x$ of interest along the critical line, and with the use of a test function. And, as mentioned in the Introduction, it is only for a restricted class of test functions --- those whose Fourier transform have support on $|k| < 2 \pi$ --- that rigorous analysis of the correlations has been possible. Even then the theorems obtained are statements about the leading asymptotic form only.

The discussion in the Introduction mentioned that a non-rigorous, but predictive analysis based on an analogy with a chaotic quantum system, together with the use of a conjecture of Hardy and Littlewood for the pair correlation function of the primes, allows for further progress \cite{BK96b}. One consequence is that (\ref{S}) can be derived as the two-point correlation function without the assumption of a restricted class of test functions. Again, this is a statement about the limiting asymptotic form. For present purposes, the most relevant feature of the results of \cite{BK96b} is that they allow for the determination of the leading correction term to the limiting asymptotic form. The necessary working is given in \cite{BBLM06}, where it was shown that at position 
${1 \over 2} + i E$, with $E \gg 1$,
along the critical line, with the local density normalised to unity, and with $\bar{\rho}$ given by the leading term in (\ref{1A})
the smoothed two-point correlation has the large $E$ expansion
\begin{equation}\label{RSc}
\rho_{(2)}^{\rm R}(E,E+s) = 1 - \Big ( {\sin \pi s \over \pi s} \Big )^2 - { \Lambda
\over \pi^2 \bar{\rho}^2 } \sin^2 \pi s -
{Q \over 2 \pi^2 \bar{\rho}^3} s \sin 2 \pi s +
{\rm O} \Big ( {1 \over \bar{\rho}^4 } \Big ).
\end{equation}
The constants $\Lambda$ and $Q$ can be expressed in terms of convergent expressions involving primes, which 
when evaluated give the numerical values $\Lambda = 1.57314...$ and $Q/\Lambda = 1.4720...$. Moreover, with
\begin{align}
\nonumber \alpha = 1 + {C \over \log(E/2 \pi)}, \qquad C = {Q \over \Lambda},
\end{align}
it was observed that (\ref{RSc}) can be rewritten
\begin{equation}\label{RSc1}
\rho_{(2)}^{\rm R}(x,x+s) = 1 - \Big ( {\sin \pi s \over \pi s} \Big )^2 - { \Lambda
\over \pi^2 \bar{\rho}^2 } \sin^2 (\pi \alpha s) + {\rm O} \Big ( {1 \over \bar{\rho}^4 } \Big ).
\end{equation}
Comparison with (\ref{Sc}) shows that the correction terms agree subject to relating the matrix size $N$ and the distance along the critical line $E$ according to \cite{BBLM06}
\begin{equation}\label{e:Neff}
N = {1 \over \sqrt{12 \Lambda} } \log \Big ( {E \over 2 \pi} \Big ),
\end{equation}
provided too that the scaled distance $s$ is further rescaled 
\begin{equation}\label{e:alphas}
s \mapsto \alpha s.
\end{equation}

In the theory of point processes (see e.g.~\cite{IPSS08}) a thinning operation, whereby each member is deleted independently with probability $(1 - \xi)$, is used to create a family of point processes from a single parent process.
The effect on the corresponding $k$-point correlation $\rho_{(k)}(x_1,\dots,x_k;\xi)$ is very simple --- the
$\xi$ dependence scales to give
\begin{equation}\label{2.8a}
\rho_{(k)}(x_1,\dots,x_k;\xi) = \xi^k \rho_{(k)}(x_1,\dots,x_k;1).
\end{equation}
Thus introducing the variables $y_j = \xi x_j$ ($j=1,2,\dots$) one sees that
\begin{align}
\label{e:rhoxi} \rho_{(k)}(x_1,\dots,x_k;\xi) dx_1 \cdots dx_k = \rho_{(k)}(y_1/\xi,\dots,y_k/\xi;1) dy_1 \cdots dy_k.
\end{align}
The variables $\{y_j\}$ correspond to a rescaling so that the density of the original point process remains unchanged, assuming  translational invariance. 

\subsection{Numerical results}

\begin{figure}[ht]
 \includegraphics[scale=0.66]{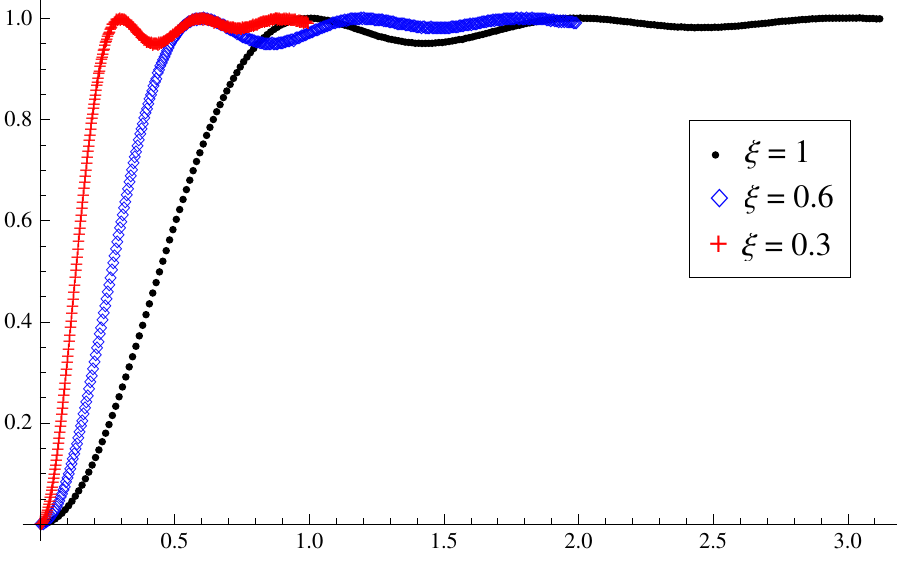} \quad \includegraphics[scale=0.66]{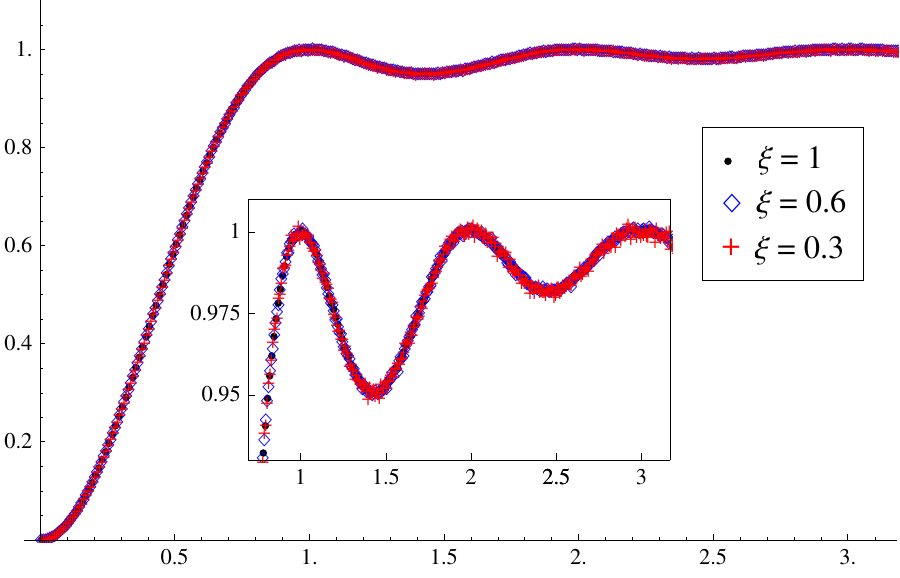} 
 \caption{\label{f:2pt}On the left is a plot of the raw 2-pt correlation function in the Riemann zero data, where each point has been deleted with probability $(1- \xi)$. The figure on the right contains the same data, which has been rescaled according to \eqref{e:rhoxi}, showing the self-similarity property for varying $\xi$.}
\end{figure}

To determine the two-point correlation $\rho_{(2)} (y_1/\xi, y_2/\xi;1)$ empirically from Odlyzko's data set we must first rescale the data by \eqref{1A} so that the rescaled local mean density is unity; we then delete each zero with probability $1- \xi$. Next we sample this sequence by empirically computing the zero density of the sub-system formed by each zero in turn as the left boundary, and a fixed number (say 50) of its neighbours.  Figure \ref{f:2pt} displays the results of this empirical determination of \eqref{e:rhoxi} in both the variables $\{ x_j \}$ (on the left) and $\{ y_j \}$ (on the right). To the eye, the resulting graphical forms are identical to the leading order random matrix prediction \eqref{S}.

\begin{figure}[ht]
 \includegraphics[scale=0.38]{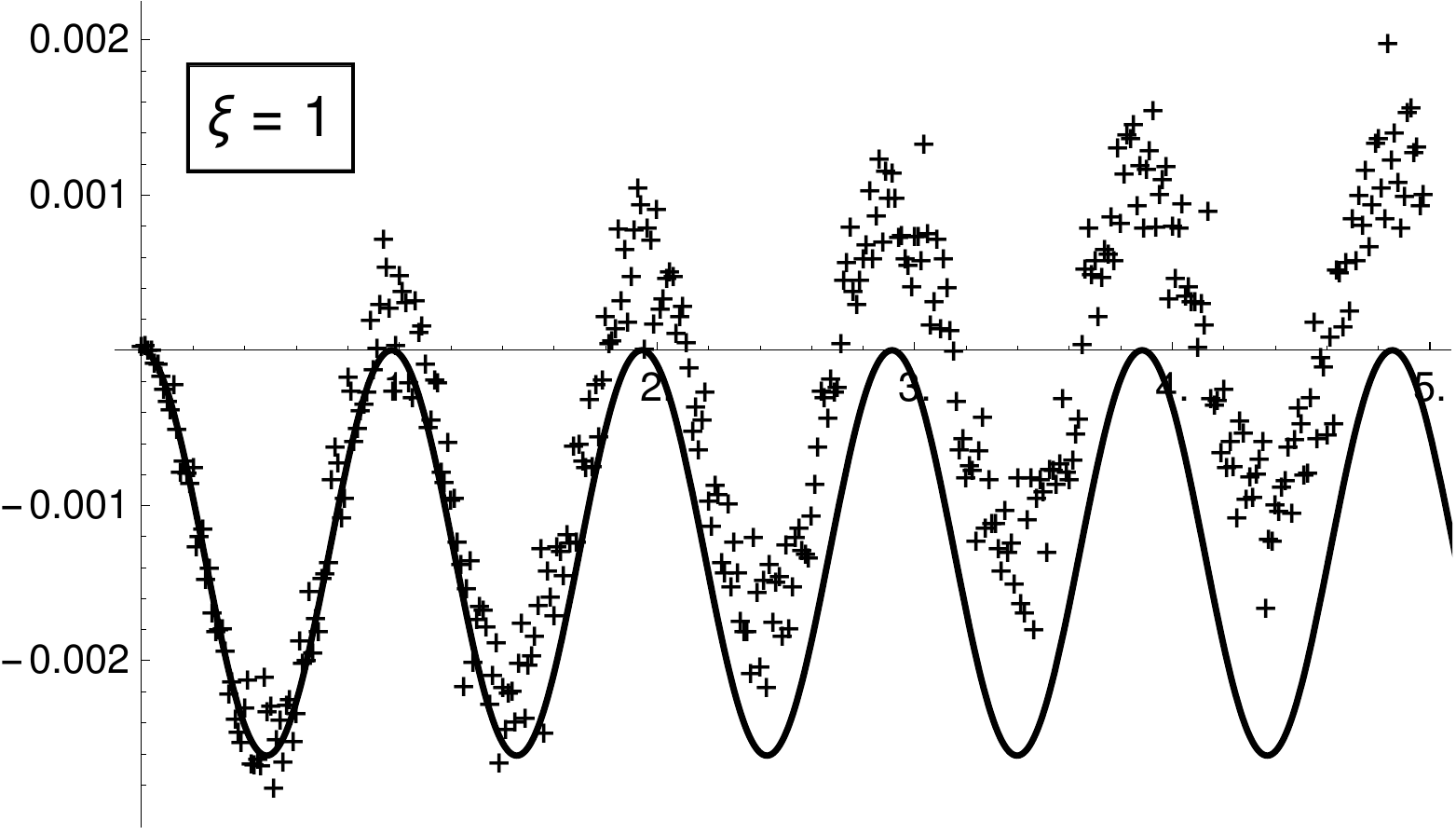} \quad \includegraphics[scale=0.38]{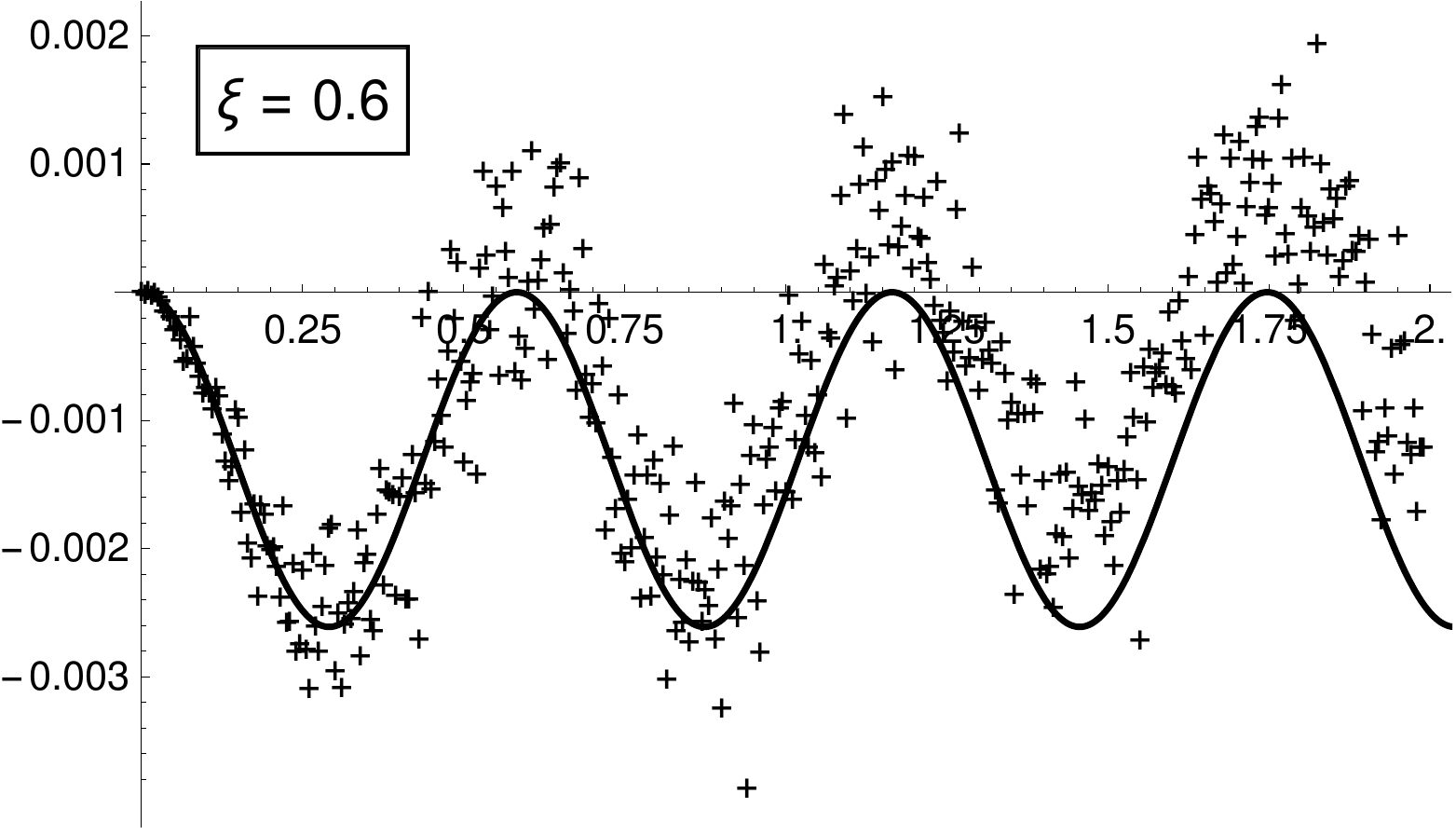} 
 \caption{\label{f:2ptcomp}Comparison of the $1/\bar{\rho}^2$ term in \eqref{RSc1} [solid curve] (with $s\to s/ \xi$) against \eqref{e:rho2Rminus} [crosses] (again with $s\to s/\xi$).}
\end{figure}

Following \cite{BBLM06} our main point of interest is in the functional form that results after subtracting the leading random matrix prediction from the empirical two-point function for Odlyzko's data set, 
\begin{align}
\label{e:rho2Rminus} \rho_{(2)}^R (E, E+s) - \left( 1 -  \Big ( {\sin \pi s \over \pi s} \Big )^2 \right).
\end{align}
The theory of \cite{BBLM06}, as reproduced in \eqref{RSc1} above, predicts that this quantity is, to leading order in $1/ \bar{\rho}$, itself of random matrix origin. As a test of its validity, Figure \ref{f:2ptcomp} displays a comparison of the empirical data and the theoretical predictions. Very good agreement is found for distances up to approximately one and a half times the average spacing between zeros, and furthermore the period of the oscillations are well matched for all distances
in the display. On the other hand, as the distances increase, there is a systematic 
discrepancy at a quantitative level. Anticipation of the expanded form \eqref{RSc1} not being in agreement 
for increasing distances comes when one graphically compares the corrections in \eqref{RSc} and \eqref{RSc1} --- recalling, in particular, that theoretically \eqref{RSc} and \eqref{RSc1} agree to leading order. Thus one finds that their mismatch increases with increasing distance. On the other hand, the functional form from \cite{BK96b} used to obtain \eqref{RSc} can itself be numerically calculated. Its graphical form has for $\xi =1$ been compared against the two-point correlation of the same sequence of Riemann zeros as used here \cite[Figures 12--17]{Bogo2006} or \cite[Figures 3--7]{Bo07} and quantitative agreement is found for all distances displayed.

\section{Spacing distributions}\label{s:space}
\subsection{Fredholm determinant formulae}
Although the $k$-point correlations cannot be empirically computed from the data beyond the case $k=2$, certain functionals of the correlations can be so computed. A specific example is the $n$\textit{-th nearest neighbour spacing distribution} $p(n; s)$, $n=0,1,2,\dots$. In terms of the correlations, and assuming a translationally invariant system with constant density $\rho$, we have in the case $n=0$,
\begin{align}
\label{e:3.0} p(0;s) = \frac{\rho_{(2)}(s,0)} {\rho} + \frac1{\rho} \sum_{j=1}^\infty {(-1)^j \over j!} \int_0^s dx_1 \cdots \int_0^s dx_j \, \rho_{(j+2)} (s,0, x_1, \dots,x_j),
\end{align}
and similar formulae for general $n$; see e.g.~\cite[\S 8.1 \& \S9.1]{Fo10}.

The calculation of the functional form of the scaled limit of $p(0; s)$ for $N\times N$ unitary matrices $U(N)$ chosen with Haar measure, or equivalently bulk scaled GUE matrices, is a celebrated problem in random matrix theory. First note that a Fredholm determinant can be expanded as a series involving multiple integrals with integrands given as determinants of the integral kernel. With this expansion one can show from the form of the scaled $k$-point correlation functions
(\ref{S1}) that
\begin{equation}\label{K}
p(0;s) = {d^2 \over d s^2} \det ( \mathbb I - \mathbb K_s),
\end{equation}
where $ \mathbb K_s$ is the integral operator supported on $(0,s)$ with kernel
\begin{align}
\nonumber K(x,y) = {\sin \pi (x - y) \over \pi (x - y)},
\end{align}
noting that the constant density $\rho$ is unity. The required working can be found, for example, in \cite[\S9.1]{Fo10}.
This Fredholm determinant formula is equivalent to the expression
$$
{d^2 \over ds^2} \prod_{l=0}^\infty (1  - \lambda_l(s)),
$$
where $1 > \lambda_1(s) > \lambda_2(s) > \cdots > 0$ are the eigenvalues of $\mathbb K_s$.
Such a  result, for the bulk scaled limit of GOE matrices, was first obtained by Gaudin \cite{Ga61}, who also provided a computational scheme by relating the eigenvalues $\{\lambda_l(s) \}$ to the eigenvalues of the second order differential operator having the prolate spheroidal functions as eigenfunctions (see e.g.~\cite[\S 9.6.1]{Fo10}). Only recently  has it been shown that there are numerical schemes based directly on (\ref{K}) 
that exhibit exponentially fast convergence to the limiting value \cite{Bo08,Bo09}.

Nearly two decades after the work of Gaudin, the Kyoto school of Jimbo et al.~\cite{JMMS80}
expressed (\ref{K}) as the so-called $\tau$\textit{-function} for a particular Painlev\'e V system. Explicitly, it was shown that
\begin{equation}\label{KS}
\det ( \mathbb I - \xi \mathbb K_s) = \exp \int_0^{\pi s} {\sigma^{(0)} (t;\xi) \over t} \, dt,
\end{equation}
where $\sigma$ satisfies the differential equation
\begin{equation}\label{d1}
(t \sigma'')^2 + 4 (t \sigma' - \sigma)(t \sigma' - \sigma + (\sigma')^2) = 0
\end{equation}
with small $t$ boundary conditions 
\begin{equation}\label{d2}
\sigma^{(0)}(t;\xi) = - {\xi \over \pi} t - {\xi^2 \over \pi^2} t^2 + {\rm O}(t^3).
\end{equation}
Note that the parameter $\xi$ introduced in (\ref{KS}) only enters in the characterisation through the boundary condition.
For background theory relating to the Painlev\'e equations as they occur in
random matrix theory we refer to~\cite[Ch.~8]{Fo10}.

Although (\ref{K}) only requires the case $\xi = 1$, the Fredholm expansion formula used to derive (\ref{K}) generalises to read
\begin{align}
\label{e:det1xiK} \det ( \mathbb I - \xi \mathbb K_s)  = \sum_{j=0}^\infty {(-\xi)^j \over j!} \int_0^s dx_1 \cdots \int_0^s dx_j \, \rho_{(j)} (x_1, \dots, x_j),
\end{align}
(cf. eq. \eqref{e:3.0}). In the case $\xi=1$, this formula, via the inclusion/exclusion principle, shows $\det (1- \mathbb{K}_s)$ can be interpreted as the probability, $E(0; s)$, say, that there are no zeros in an interval of length $s$ of the original data set. Thus we have
\begin{align}\label{Ep}
E(0; s) = \det ( \mathbb I - \mathbb K_s).
\end{align}
In the case of $0< \xi< 1$, \eqref{2.8a} together with the inclusion/exclusion principle shows \eqref{e:det1xiK} is equal to the probability that there are no zeros in an interval of length $s$ in the data set corresponding to a thinning of the original data set by deleting each zero with probability $(1- \xi)$ --- this is the procedure alluded to in the Introduction. With $p(0;s; \xi)$ denoting the corresponding distribution of the nearest neighbour spacing, \eqref{K} generalises to
\begin{align}
 \label{e:ps0xi} p(0;s; \xi)= \frac1{\xi} \frac{d^2} {ds^2} \det \left( \mathbb{I} -\xi \mathbb{K}_s \right),
\end{align}
where the the constant density $\rho =\xi$ in this decimated system. We remark that the notation $p((0,s);\xi)$ has been used in \cite[\S 8.1]{Fo10} to denote the generating function for the conditioned spacing distributions $\{p(n;s)\}_{n=0,1,\dots}$, corresponding to the conditioned gap probabilities in the next paragraph. These quantities are related by
\begin{align}\label{3.8a}
\frac1{\xi} \; p(0;s;\xi) = p((0,s);\xi).
\end{align}

As just mentioned, another interpretation shows itself via the formula (see e.g.~\cite[eq.~(8.1) together with (9.15)]{Fo10})
\begin{equation}\label{Eq}
E(m; s) = {(-1)^m \over m!} {\partial^m \over \partial \xi^m} \det ( \mathbb I - \xi \mathbb K_s) \Big |_{\xi = 1},
\end{equation}
where $E(m ;s)$ denotes the probability of an interval $s$ in the original data set containing exactly $m$ zeros. This is referred to as a \textit{(conditioned) gap probability}. Note that (\ref{Eq}) reduces to (\ref{Ep}) in the case $m=0$.

Both the expressions (\ref{K}) and (\ref{KS}) have analogues for the eigenvalues of $U(N)$ itself, rather than their scaled limit. Let $\mathbb{K}_s^N$ denote the integral operator supported on $(0, s)$ with kernel $K_N (x, y)$ as specified by \eqref{2.0}. The analogue of \eqref{e:ps0xi} is then the structurally identical formula
\begin{equation}\label{2.13a}
p^N (0; s; \xi) = \frac1{\xi}  {d^2 \over d s^2} \det ( \mathbb I - \xi \mathbb K_s^N).
\end{equation}
Here $p^N (0; s; \xi)$ denotes the nearest neighbour spacing distribution for matrices from $U (N)$ chosen with Haar measure, with $s$ the difference in neighbouring angles. The $\tau$-function formula (\ref{KS}) generalises to
\cite[eq.~(1.33)]{FW06}
\begin{equation}\label{2.13b}
\det ( \mathbb I - \xi \mathbb K_s^N) = \exp \Big (- \int_0^{\pi s/N} U(\cot \phi;\xi) \, d \phi \Big ),
\end{equation}
where $U$ satisfies the $\tilde{\sigma}$PVI equation
(see e.g.~\cite[eq. (8.21)]{Fo10})
\begin{equation}\label{q5}
u'((1+s^2)u'')^2 + 4 (u'(u-su') + i v_1 v_2 v_3 v_4)^2 + 4 \prod_{k=1}^4 ( u' + v_k^2) = 0,
\end{equation}
with $u(s) = U (s; \xi)$ and parameters
\begin{align}
\nonumber v_1 = v_2 = v_3 = 0, \quad v_4 = N,
\end{align}
subject to the boundary condition $u(s) \sim \xi N/\pi$ as $s \to \infty$.

\subsection{Finite $N$ correction for the nearest neighbour spacing}

Our interest is in the leading correction term to the large $N$ form of (\ref{2.13a}).
Since, from (\ref{2.0}),
$$
K_N(x,y) = {\sin \pi (x - y) \over \pi (x - y)} + {\pi (x - y) \over 6 N^2} \sin \pi (x - y) + {\rm O} \Big ( {1 \over N^4} \Big ),
$$
we see that this correction term is of order $1/N^2$. Specifically
\begin{align}\label{Tq0}
\det ( \mathbb I - \xi \mathbb K_s^N) = \det ( \mathbb I - \xi \mathbb K_s) \left[ 1-  \frac{\xi}{N^2} {\rm Tr} \, \Big (  ( \mathbb I - \xi \mathbb K_s)^{-1} L_s \Big )\right] + {\rm O} \Big ( {1 \over N^4} \Big ),
\end{align}
where $L_s$ is the integral operator on $(0,s)$ with kernel $(\pi(x-y)/6) \sin \pi (x - y)$.
We remark that with $R(x,y)$ denoting the resolvent kernel --- that is the kernel supported on $(0,s)$ of the integral operator $\xi \mathbb K_s ( \mathbb I - \xi \mathbb K_s)^{-1}$ --- straightforward manipulation shows
\begin{align}
\nonumber {\rm Tr} \, \Big (  ( \mathbb I - \xi \mathbb K_s)^{-1} L_s \Big ) = {\pi \over 6} \int_0^s dx \int_0^s dy \, R(x,y) (y - x) \sin \pi (y - x).
\end{align}

The expansion (\ref{Tq0}) tells us that in relation to the representation (\ref{2.13b}) we should change variables and write
\begin{equation}\label{2.13c}
- {1 \over N} U(\cot X/N;\xi) =
{\sigma^{(0)}(X) \over X} + {1 \over N^2} {\sigma^{(1)}(X) \over X}  + {\rm O} \Big ( {1 \over N^4} \Big ).
\end{equation}
This implies the expansion
\begin{equation}\label{2.13d}
\det ( \mathbb I - \xi \mathbb K_s^N) = 
\exp \left( \int_0^{\pi s} {\sigma^{(0)}(X) \over X} \, dX \right)
\left( 1 + {1 \over N^2} \int_0^{\pi s} {\sigma^{(1)}(X) \over X} \, dX + {\rm O} \Big ( {1 \over N^4} \Big ) \right ).
\end{equation}
Both $\sigma^{(0)}$ and $\sigma^{(1)}$ also depend on $\xi$ (which, as mentioned below \eqref{d2}, enters through the boundary conditions), but for notational convenience this has been
suppressed. As noted above, $\sigma^{(0)}$  satisfies the particular Painlev\'e V equation in sigma form
(\ref{d1}) with boundary condition (\ref{d2}). 
By changing variables $s = \cot X/N$ in (\ref{q5}) and introducing the expansion (\ref{2.13c}), the equation (\ref{d1}) 
can be reproduced (a fact already known from \cite{FW04}), and moreover
a linear differential equation for $\sigma^{(1)}(X)$ with coefficients involving $\sigma^{(0)}(X)$, can be obtained.

\begin{prop}\label{P1}
With $\sigma^{(0)}(X)$, $\sigma^{(1)}(X)$ related to $U(s; \xi)$ in (\ref{2.13b}) by (\ref{2.13c}) we have that $\sigma^{(0)} (X)$ satisfies the particular Painlev\'e V equation in sigma form (\ref{d1}) with boundary condition (\ref{d2}), while $\sigma^{(1)}(X)$ satisfies the second order, linear differential equation
\begin{equation}\label{Sig2}
A(s) y''(s) + B(s) y'(s) + C(s) y(s) = D(s),
\end{equation}
where, with $\sigma(s) = \sigma^{(0)}(s)$,
\begin{align}
\nonumber A(s) & = 2 s^2 \sigma''(s),\\
\nonumber B(s) & = - 8  \sigma'(s) \sigma(s) + 12 s (\sigma'(s))^2 + 8s \Big( s \sigma'(s) - \sigma( s) \Big),\\
\nonumber C(s) & = - 4 (\sigma'(s))^2 - 8 \Big( s \sigma'(s) - \sigma(s) \Big), \\
\nonumber D(s) & = -\frac{4} {3} s^2 \sigma''(s) \left( \sigma(s) - s  \sigma'(s) - { s^2 \over 2} \sigma''(s) \right) \\
\nonumber & - \frac{4} {3} (s  \sigma'(s) - \sigma(s)  )\Big( 3 (\sigma(s))^2 + 2s \sigma(s) \big( s- \sigma' (s) \big)- 2s^2 \sigma' (s) \big( s+ \sigma' (s) \big) \Big).
\end{align}
The corresponding $s \to 0^+$ boundary condition is 
\begin{align}\label{bcs}
\sigma^{(1)}(s) & = - \Big ( s^4 {\xi^2 \over 9 \pi^2} + s^5 {5 \xi^3 \over 36 \pi^3} + {\rm O}(s^6) \Big ).
\end{align}
\end{prop}

\noindent
{\it Proof.} \quad
We see there are three main terms in (\ref{q5}). Changing variables $s = \cot X/N$ and introducing the expansion (\ref{2.13c})
in the first gives
\begin{align}
\nonumber &\Big ( (1 + s^2) U''(s; \xi) \Big )^2  = N^2 \left\{ X^2 \left( \frac{d^2 \sigma^{(0)} (X)} {dX^2} \right)^2 \right\} \\
\nonumber &+ 4X^2 \frac{d^2 \sigma^{(0)} (X)} {dX^2} \left( \frac1{2} \frac{d^2 \sigma^{(1)} (X)} {dX^2} - \frac{X}{3} \frac{d \sigma^{(0)} (X)} {dX} + \frac{\sigma^{(0)}(X)} {3} - \frac{X^2} {6} \frac{d^2 \sigma^{(0)} (X)} {dX^2} \right),
\end{align}
up to terms of order $1/N^4$. Similarly expanding the other two main terms and then equating terms of order $N^2$ gives
(\ref{d1}), while equating terms independent of $N$ gives (\ref{Sig2}). 

In relation to the boundary conditions, for $U(s; \xi)$ in (\ref{2.13b}) we know from \cite[eq.~(8.78)]{Fo10}, further extended to the next order,  that for large $s$
\begin{align}
\nonumber U (s; \xi) = & c+ \frac{c^2} {s} +\frac{c^3} {s^2} + c^2 \frac{9c^2 -N^2 -2} {9 s^3} + c^3 \frac{36c^2- 5N^2 -19}{36 s^4}\\
\nonumber & + c^2 \frac{46-375c^2+450c^4+N^2 (40-75c^2) +4N^4} {450 s^5} + {\rm O} (s^{-6}),
\end{align}
where $ c =  \frac{\xi N} {\pi}$. Substituting in (\ref{2.13c}) and performing appropriate additional expansions we read off
(\ref{bcs}). \hfill $\square$

\medskip
While we know of no other analytic formulae for the correction term to the spacing distribution for random matrix
ensembles in the bulk, at an edge --- specifically the soft edge --- this task seems to have been first taken up by \cite{Ch06}.
Aspects of this same problem at the hard edge are discussed in the recent works
\cite{EGP14,Bo15,PS15}.

Let $f^N (x) [N^{-p}]$ denote the coefficient of $N^{-p}$ in the large $N$ expansion of $f^N (x)$. Then we have from \eqref{2.13d} that
\begin{equation}\label{C1}
\det ( \mathbb I - \xi \mathbb K_s^N) [N^{-2}] = \Big (  \int_0^{\pi s} {\sigma^{(1)}(X) \over X} \, dX  \Big ) \exp \Big ( \int_0^{\pi s} {\sigma^{(0)}(X) \over X} \, dX \Big ).
\end{equation}
One use of Proposition \ref{P1} is to be able to deduce the explicit form of the small and large distance expansions of \eqref{C1}.
We will 
consider first the small $s$ expansion. 

\begin{cor}
We have
{\small\begin{align}
\nonumber &\sigma^{(0)}(s) = -\frac{\xi  s }{\pi } -\frac{ \xi^2 s^2}{\pi ^2} -\frac{\xi^3 s^3}{\pi ^3}  + \frac{1}{24} \left(\frac{8\xi^2 }{3 \pi ^2}-\frac{24 \xi^4}{\pi ^4}\right) s^4 + \left(\frac{5 \xi^3}{36 \pi ^3}-\frac{\xi^5}{\pi ^5}\right) s^5\\
&+ \left(-\frac{\xi^6}{\pi ^6}+\frac{\xi^4}{6 \pi ^4}-\frac{2 \xi^2}{225 \pi ^2}\right) s^6 + \left(-\frac{\xi^7}{\pi ^7}+\frac{7 \xi^5}{36 \pi ^5}-\frac{7 \xi^3}{675 \pi ^3}\right) s^7 \nonumber \\
  &  + \left(-\frac{\xi^8}{\pi ^8} +\frac{2 \xi^6}{9 \pi ^6}-\frac{121 \xi^4}{8100 \pi ^4}+\frac{\xi^2}{2205 \pi ^2}\right) s^8 + \left(-\frac{\xi^9}{\pi ^9}+\frac{\xi^7}{4 \pi ^7}-\frac{73 \xi^5}{3600 \pi ^5}+\frac{761 \xi^3}{1587600 \pi ^3}\right) s^9\nonumber \\
  & + \mathrm{O} \left( s^{10} \right), \label{Sd0}
  \end{align}}
{\small \begin{align}
\nonumber &\sigma^{(1)}(s) =  - {\xi^2 s^4 \over 9 \pi^2} - {5 \xi^3 s^5 \over 36 \pi^3} + {(-15 \xi^4 + 2 \pi^2 \xi^2) s^6 \over 90 \pi^4} + {7(-15 \xi^5 + 2 \xi^3 \pi^2) s^7 \over 540 \pi^5}  \\
\nonumber &-\frac{\left( 1260 \xi^6-203 \pi ^2 \xi^4+12 \pi ^4 \xi^2 \right) s^8}{5670 \pi ^6} - \frac{\left( 9450 \xi^7-1785 \pi ^2 \xi^5+83 \pi ^4 \xi^3 \right) s^9} {37800 \pi ^7}\\
&+\mathrm{O} \left( s^{10} \right), \label{Sd1}
 \end{align}}
{\small \begin{align}
\det ( \mathbb I - \xi \mathbb K^N_s) [1]  & = 1 - \xi s + { \xi^2 \pi^2  \over 36} s^4 -  {\xi^2 \pi^4 \over 675} s^6 + {\xi^2 \pi^6 \over 17640} s^8 -\frac{\xi^3 \pi^6} {291600} s^9 +\mathrm{O} \left( s^{10} \right), \label{A1} \\   
\det ( \mathbb I - \xi \mathbb K_s^N) [N^{-2}] & = - {\xi^2 \pi^2 s^4 \over 36} + {\xi^2 \pi^4 s^6 \over 270} - {\xi^2 \pi^6 s^8 \over 3780} + \frac{\xi^3 \pi^6 s^9} {48600} + \mathrm{O} \left( s^{10} \right), \label{Sd2}
 \end{align}}
\begin{align}
{2 \pi \over N} p^N(0; 2 \pi s/N ;\xi)[1] & = \frac{\xi \pi ^2 s^2} {3}-\frac{2 \xi \pi ^4 s^4}{45}+ \frac{\xi \pi ^6 s^6}{315} -\frac{\xi^2 \pi ^6 s^7}{4050} +\mathrm{O} \left( s^{8} \right), \label{P12} \\
{2 \pi \over N}  p^N(0; 2 \pi s/N ;\xi)[N^{-2}] & = -\frac{\xi \pi ^2 s^2}{3}+\frac{\xi \pi ^4 s^4}{9} -\frac{2 \xi \pi ^6 s^6}{135} +\frac{\xi^2 \pi ^6 s^7}{675} +\mathrm{O} \left( s^{8} \right).
\label{P13} \end{align}
\end{cor}

\noindent
{\it Proof.} \quad  The differential equation (\ref{d1}) admits a unique power series expansion about the origin
with the first two terms given by (\ref{d2}). Expanding to higher order and solving for the unspecified coefficients gives
(\ref{Sd0}).
Substituting the expansion in \eqref{KS} gives \eqref{A1} for the small $s$ expansion of the Fredholm determinant; this series
is also reported in  \cite[eq.~(8.114)]{Fo10}.
If we substitute (\ref{Sd0}) instead in  the differential equation (\ref{Sig2}) and make use of the boundary
condition (\ref{bcs}) we find a unique power series expansion solution is generated which upon solving for the
unspecified coefficients gives
(\ref{Sd1}). Substituting \eqref{Sd0} and \eqref{Sd1} in the order $1/ N^2$ correction term \eqref{C1}
gives (\ref{Sd2}). Recalling (\ref{2.13a}) and that we have a uniform density $\rho_{(1)} (s)= \xi$, we see (\ref{P12}) and (\ref{P13}) follow immediately from 
(\ref{A1}) and (\ref{Sd2}).
\hfill $\square$

\medskip

A consistency check can be placed on the above expansions. For this we note 
that it is also possible to use the characterisation (\ref{2.13b}) of $\det ( \mathbb I - \xi \mathbb K_s^N)$  to deduce the $N$-dependent
small $s$ expansion \cite[eq.~(8.79)]{Fo10}
\begin{eqnarray}\label{8.cue1}
 \det ( \mathbb I - \xi \mathbb K_s^N)   & = & 1 - \xi s + {(1-1/N^2)\xi^2 \pi^2 s^4 \over 36} -
{(1-1/N^2)(2-3/N^2) \over 1350} \xi^2 \pi^4 s^6 
\nonumber \\
&& \! \! + {(1-1/N^2)(1-2/N^2)(3-5/N^2) \over 52920} \xi^2  \pi^6 s^8 
  + {\rm O}(s^{9}).
\end{eqnarray}
Expanding the RHS of (\ref{8.cue1}) for large $N$ we obtain agreement with (\ref{A1}) for the leading form, and agreement with (\ref{Sd2}) for the next leading order $1/N^2$ term.

A noteworthy feature of (\ref{P12}) and (\ref{P13}) is the weak dependence on the dilution parameter $\xi$, which only changes from $\xi$ to $\xi^2$ at order $s^7$. One understanding of this result relates to the interpretation of (\ref{e:ps0xi}), upon division by $\xi$, as the generating function for $\{p(n;s) \}_{n=0, 1, \dots}$; recall \eqref{3.8a}. Then, in analogy with (\ref{Eq}), $p(1;s)$ is obtained from (\ref{P12}) and (\ref{P13}) by applying $- {\partial \over \partial \xi}$ and setting $\xi = 1$. It follows that $p(1;s)$ has leading small $s$ behaviour proportional to $s^7$, which is a known result \cite[eq.~(8.115)]{Fo10}.

We turn our attention now to the behaviours for large $s$.  The asymptotics of spacing distributions for random matrix ensembles in this regime have been reviewed
in the recent work \cite{Fo14}. The case $\xi = 1$  must be distinguished  from $0 < \xi < 1$.

\begin{cor}\label{Cor3}
Suppose $0 < \xi < 1$. As $s \to \infty$ we have \cite[eq.~(1.16), with $\sigma_0 (s/2) = \sigma^{(0)} (s)$ in our notation]{MT86}
\begin{equation}\label{SS}
\sigma^{(0)}(s) = - k s + {1 \over 2} k^2 + {\rm O}\left( {1 \over s} \right), \qquad k = - \frac1{\pi} \log (1 - \xi),
\end{equation}
and thus
\begin{equation}\label{SS1}
\det ( \mathbb I - \xi \mathbb K_s^N)[1] \mathop{\sim}\limits_{s \to \infty} A(\xi )s^{k^2/2} e^{-k \pi s}.
\end{equation}
Also 
\begin{align}
\label{e:SSsig1} \sigma^{(1)}(s) =  -{ k^2 s^2 \over 6}  + {\rm O}(s)
\end{align}
and this together with (\ref{SS1}) implies
\begin{equation}\label{SS2}
\det ( \mathbb I - \xi \mathbb K_s^N) [N^{-2}] = A(\xi) s^{k^2/2} e^{-k \pi s} \left( - {k^2 \over 12} (\pi s)^2  + {\rm O}(s) \right).
\end{equation}

Suppose instead $\xi = 1$. In this case, for $s \to \infty$ we have
\begin{equation}\label{SSa}
\sigma^{(0)}(s) = - \frac{s^2} {4}  - \frac1{4} +  {\rm O} \left( {1 \over s^2} \right),
\end{equation}
which is equivalent to the expansion
\begin{align}
\label{e:det1Ks} \det (\mathbb I - \mathbb K_s^N)[1] \mathop{\sim}\limits_{s \to \infty} A(1) {e^{-(\pi s)^2/8} \over (\pi s)^{1/4}}.
\end{align}
We also have
\begin{align}
\label{SSb} \sigma^{(1)}(s) = -\frac{s^4} {48} + \frac{s^2} {48} + {\rm O}(1),
\end{align}
which when combined with (\ref{e:det1Ks}) is equivalent to
\begin{equation}\label{SS2a}
\det ( \mathbb I - \xi \mathbb K_s^N) [N^{-2}] = A(1) {e^{-(\pi s)^2/8} \over (\pi s)^{1/4}}  \left( - {(\pi s)^4 \over 192} + {(\pi s)^2 \over 96} + {\rm O}(s) \right).
\end{equation}

Thus, for large $s$
\begin{align}
 \label{SDa} {2 \pi \over N}  p^N(0; 2 \pi s/N ;\xi)[1]  &\mathop{\sim}\limits_{s\to \infty}
 \left\{ \begin{array}{lc}
 \frac{A (\xi)} {\xi} (k \pi)^2 s^{k^2/2} e^{-k \pi s}, &0 < \xi< 1,\\
A (1) \frac{\pi^{15/4}} {16} s^{7/4} e^{-(\pi s)^2/8}, &\xi =1,
 \end{array}\right.\\
 \label{SDb} {2 \pi \over N}  p^N(0; 2 \pi s/N ;\xi)[N^{-2}]  &\mathop{\sim}\limits_{s\to \infty}
 \left\{ \begin{array}{lc}
 -\frac{A (\xi)} {\xi} \frac{(k \pi)^4} {12} s^{k^2/2+2} e^{-k \pi s}, &0 < \xi< 1,\\
 -A (1) \frac{\pi^{31/4}} {3072} s^{23/4} e^{-(\pi s)^2/8},  &\xi =1.
 \end{array}\right.
\end{align}

\end{cor}

\noindent
{\it Proof.} \quad
As noted, the expansion (\ref{SS}) can be read off from \cite{MT86}, and this substituted in (\ref{KS}) gives (\ref{SS1}). The value of $A(\xi)$ is given in \cite[eq.~(1.14)]{BI13}, although its derivation requires different methods \cite{BB95,BW83}. Substituting (\ref{SS}) in (\ref{Sig2}) and solving for large $s$ gives (\ref{e:SSsig1}), and this together with knowledge of (\ref{SS1}) gives (\ref{SS2}).

From the definition of $k$, we see that the behaviour (\ref{SS}) breaks down when $\xi = 1$. In that case we have, instead
of (\ref{SS}), the large $s$ expansion (\ref{SSa}). 
This follows from (\ref{KS}) and knowledge of the leading two terms of the large $s$ form of $\det (\mathbb I - \mathbb K_s)$
\cite{DIZ97, VV10} as given by (\ref{e:det1Ks}).
Substituting (\ref{SSa}) in (\ref{Sig2}) and solving for large $s$ gives (\ref{SSb}).
Then substituting this in (\ref{C1}) and using (\ref{e:det1Ks}), we see that the large $s$ form of the $1/N^2$ correction term \eqref{C1} for $\xi =1$ is given by
\eqref{SS2a}.

To obtain the behaviour of the spacing distribution we use \eqref{SS1}, \eqref{SS2}, \eqref{e:det1Ks} and \eqref{SS2a} in conjunction with \eqref{2.13a} and so deduce  (\ref{SDa}) and \eqref{SDb}. \hfill$\square$

\medskip

Numerical methods to be discussed in Section \ref{s:Num} allow the large $s$ forms of $\sigma^{(0)} (s)$ and $\sigma^{(1)} (s)$ to be compared against their computed values from the differential equations; see Figures \ref{f:pwfnxi100}--\ref{f:diffs60}. A prominent feature seen upon differentiating the transcendents is that for $0 < \xi <1$ higher order terms in the large $s$ expansion are oscillatory, as known from the analytic work of McCoy and Tang \cite{MT86} in the case of $\sigma^{(0)} (s)$.
 Since there is no such effect for $\xi =1$, we display only the transcendent and its asymptotic form, and not the derivative. The asymptotic form in this case is functionally distinct from that for $0< \xi <1$ --- note that the quantity $k$ in the latter actually diverges at $\xi=1$.

\begin{figure}[ht]
 \includegraphics[scale=0.66]{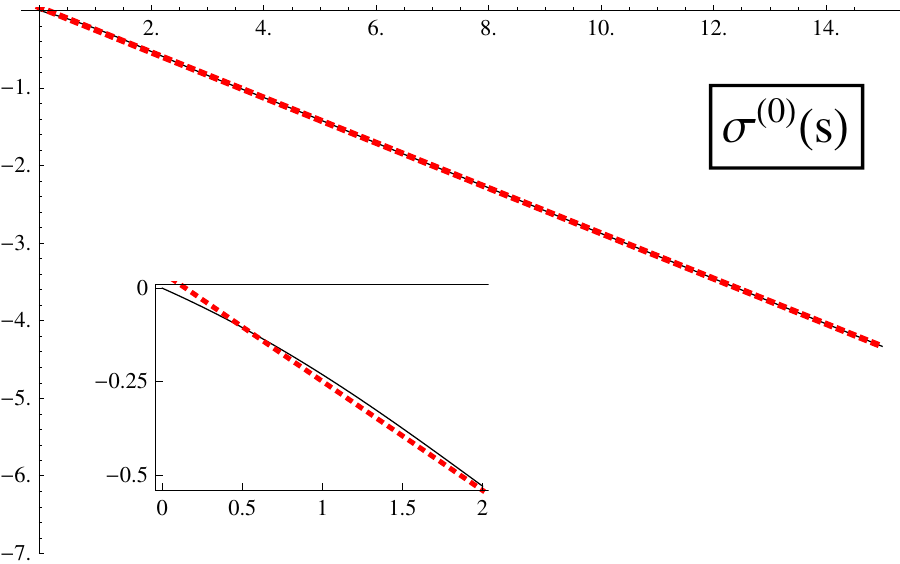} \quad \includegraphics[scale=0.66]{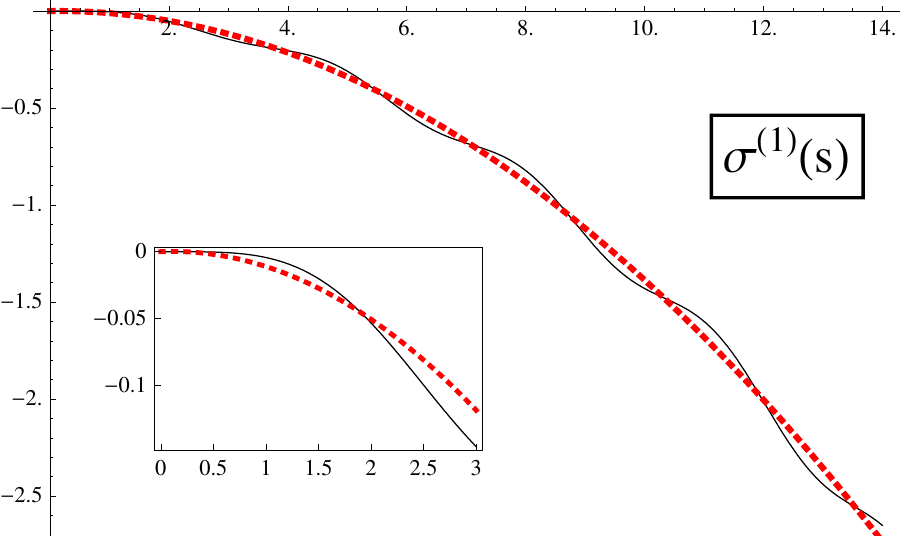}
 \caption{\label{f:pwfnxi60}$\boldsymbol{\xi=0.6}$. The black curves are the solutions of the DEs \eqref{d1} [left] and \eqref{Sig2} [right] using a sequence of Taylor series of degree $12$. The dashed red curves are the asymptotic approximations \eqref{SS} [left] and \eqref{e:SSsig1} [right]. The inset is the same plot zoomed-in on the origin.}
\end{figure}

\begin{figure}[ht]
 \includegraphics[scale=0.66]{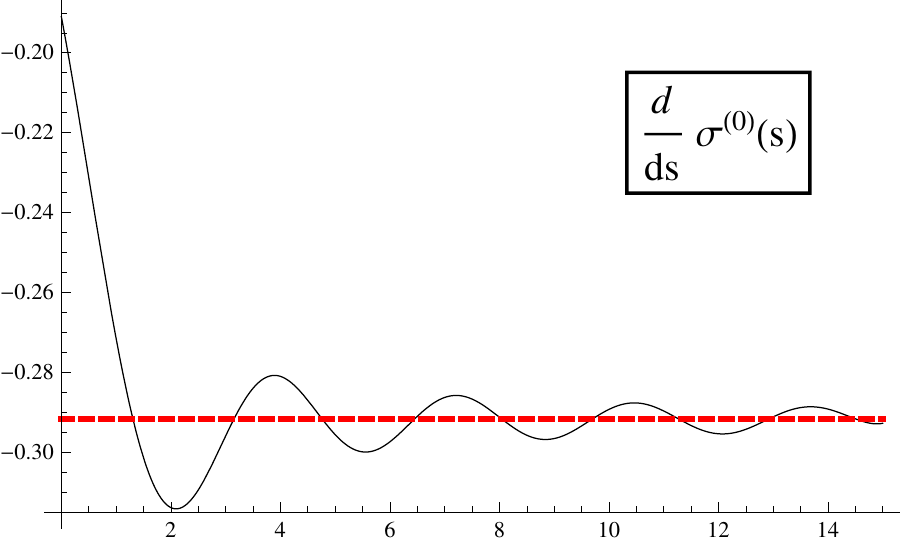} \quad \includegraphics[scale=0.66]{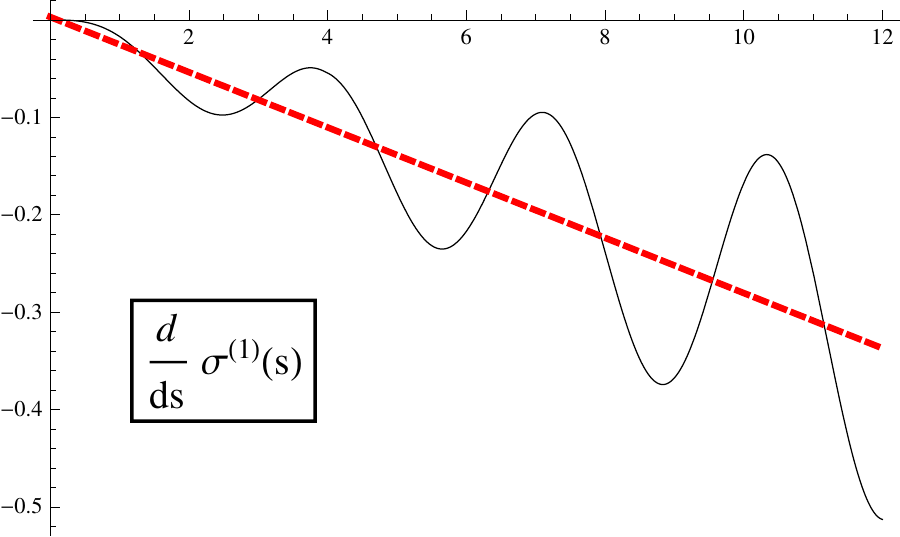}
 \caption{\label{f:diffs60}$\boldsymbol{\xi=0.6}$. These are the derivatives w.r.t. $s$ of the corresponding curves in Figure \ref{f:pwfnxi60}.}
\end{figure}

\begin{figure}[ht]
 \includegraphics[scale=0.66]{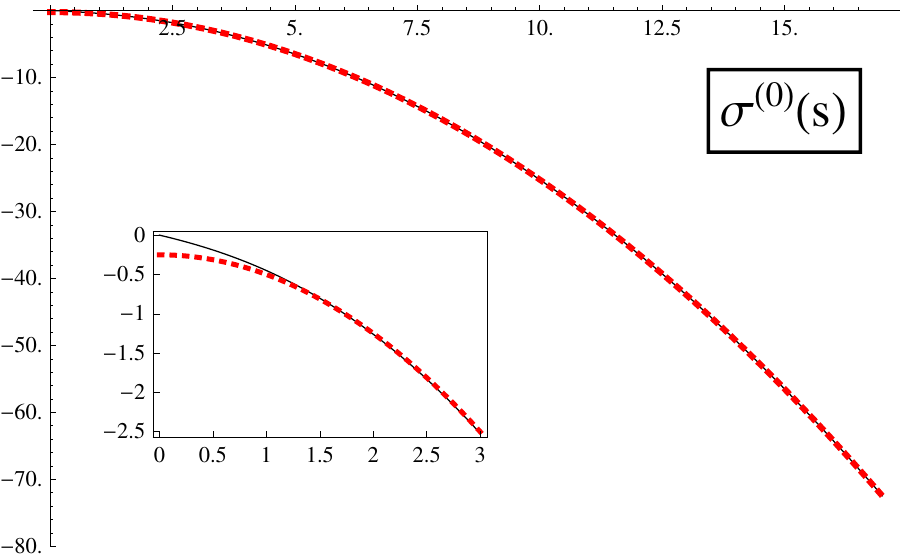} \quad \includegraphics[scale=0.66]{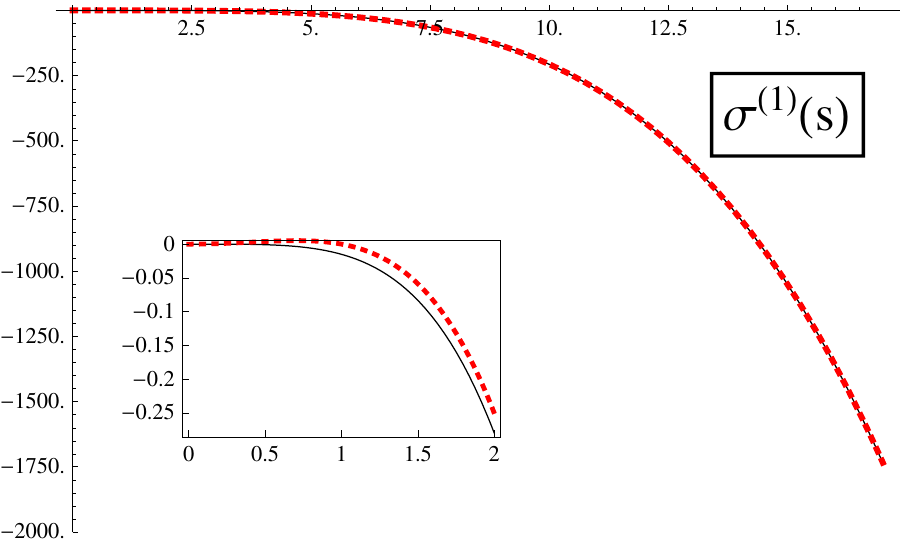}
 \caption{\label{f:pwfnxi100}$\boldsymbol{\xi=1}$. The black curves are the solutions of the DEs \eqref{d1} [left] and \eqref{Sig2} [right] using a sequence of Taylor series of degree $35$. The dashed red curves are the asymptotic approximations \eqref{SSa} [left] and \eqref{SSb} [right]. The inset is the same plot zoomed-in on the origin.}
\end{figure}

A variation on the process of deleting each zero with the (constant) probability $1-\xi$ is to delete with probability $1- \xi (x)$, with $x=0$ corresponding to some pre-determined origin in the data. Since upon the scaling \eqref{1A} the original data set is translationally invariant,  an ensemble can be generated by varying the origin. The corresponding $k$-point correlation, $\rho_{(k)} (x_1, \dots, x_k; \xi (x))$ say, has the factorisation property 
\begin{align}
\nonumber \rho_{(k)} (x_1, \dots, x_k; \xi (x)) = \left( \prod_{l=1}^k \xi (x_l) \right) \rho_{(k)} (x_1, \dots, x_k; 1)
\end{align}
(cf. \eqref{2.8a}). Thus in this circumstance the correlation kernel in \eqref{S1} should be replaced by
\begin{align}
\label{e:SV} \big( \xi (x_j) \xi (x_l) \big)^{1 /2} \frac{\sin \pi (x_j -x_l)} {\pi (x_j- x_l)}.
\end{align}
Denoting the integral operator supported on $(0, s)$ with kernel \eqref{e:SV} by $\mathbb{K}_s [\xi]$, the probability that an interval of size $s$ from the origin has no zeros is equal to $\det (1- \mathbb{K}_s [\xi])$. Subject to $\xi(x)$ having the functional form $\xi (x/s)= v(y)$, $y= x/s$ with $v(y)$ analytic on $[0, 1]$, then (for large $s$) the expansion \eqref{SS1} generalises to \cite[Theorem 2.1]{KitaKozlMailSlavTerr09}
\begin{align}
\nonumber \det (1- \mathbb{K}_s [\xi]) \sim C[\xi] s^{(a_0^2 +a_1^2)/4} e^{-\pi k_v s},
\end{align}
where
\begin{align}
 \nonumber k_v= -\frac1{\pi} \int_0^1 \log \big(1- v(x) \big) dx, \qquad a_y = -\frac1{\pi} \log \big(1-v (y) \big),
\end{align}
and for a certain explicit $C[\xi] $. Unfortunately the requirement that $\xi$ scale with $s$ prohibits probing this
process in the Riemann zeros data. This is similarly true of the choice $\xi = 1 - e^{-2 \kappa s}$ recently studied
for large $s$  in \cite{BDIK15}.

\subsection{Alternative characterisation of finite $N$ correction for nearest neighbour spacing}
We see from (\ref{2.13a}) that to obtain the order $1/N^2$ correction to the limiting spacing distribution from knowledge of the $1/N^2$ correction to the gap probability, the second derivative with respect to $s$ of the latter must be taken. In fact, as first shown in \cite{FW00} and further refined in \cite{FW04}, the second derivative can be incorporated within the Painlev\'e theory. Specifically, we have from \cite{FW04} (again recalling \eqref{3.8a}) that
\begin{align}
\nonumber p(0;s; \xi) = {\xi \pi^2 s^2 \over 3} \exp \int_0^{2\pi s} u^{(0)}(t;\xi) \, {dt \over t} ,
\end{align}
where $u^{(0)}$ satisfies the particular, modified $\sigma$PV equation
\begin{align}\label{m1a}
\Big( s u''(s) \Big)^2 + \Big( s u'(s) - u(s) \Big) \Big(s u'(s) - u(s) -4 + 4(u'(s))^2 \Big) - 16 \Big( u'(s) \Big)^2 = 0
\end{align}
subject to the small $s$ boundary condition
\begin{equation}\label{m2}
u^{(0)}(s;\xi) = - {1 \over 15} s^2 + {\rm O}(s^4) -{\xi  \over 8640 \pi} \Big (s^5 + {\rm O}(s^7) \Big ).
\end{equation}
Moreover, the result \cite[eq.~(5.16)]{FW04} also tells us that for finite $N$
\begin{align}
\nonumber {2 \pi \over N} p^N(0; 2 \pi s/N ; \xi) = {\xi \over 3} (N^2 - 1) \sin^2{ \pi  s \over N} \exp \left(- \int_0^{\pi s/N} V(\cot \phi;\xi) \, d \phi \right),
\end{align}
where $V$ satisfies the $\tilde{\sigma}$PVI equation (\ref{q5}) with $u(s) = V(s; \xi)$ and parameters
\begin{equation}\label{m4}
v_1 = v_4 = 0, \quad v_2 = - N, \quad v_3 = - 2.
\end{equation}

Analogous to (\ref{2.13c}), in keeping with the correction to the large $N$ form being of order $1/N^2$,
we expand
\begin{align}
\nonumber - {1 \over N} V(\cot X/(2N);\xi) ={u^{(0)}(X) \over X} + {1 \over N^2} {u^{(1)}(X) \over X}  + {\rm O} \left( {1 \over N^4} \right),
\end{align}
so that
\begin{align}
\nonumber {2 \pi \over N} p^N(0; 2 \pi s/N ; \xi)&= \frac{\xi \pi^2 s^2} {3} \exp \left( \int_0^{2\pi s} {u^{(0)}(X) \over X} \, dX \right)\\
\label{3.47} &\times \left( 1- \frac{1} {N^2} - \frac{\pi^2 s^2} {3 N^2} + \frac{1} {N^2} \int_0^{2 \pi s} {u^{(1)}(X) \over X} \, dX + {\rm O} \left( {1 \over N^4} \right) \right).
\end{align}
We know that $u^{(0)}(X)$ satisfies the nonlinear equation (\ref{m1a}) subject to the
boundary condition (\ref{m2}).
Analogous to Proposition \ref{P1} it is possible to specify $u^{(1)}(X)$ as a second order linear differential equation with coefficients involving $u^{(0)} (X)$.

\begin{prop}\label{P2}
We have that $u^{(1)}(X)$ satisfies the second order, linear differential equation
\begin{equation}\label{e:u1}
\tilde{A}(s) y''(s) + \tilde{B}(s) y'(s) + \tilde{C}(s) y(s) = \tilde{D}(s),
\end{equation}
where, with $u(s) = u^{(0)}(s)$,
\begin{align}
\tilde{A} (s) & = 8 s^2 u''(s), \nonumber \\
\tilde{B} (s) & = 8\left( 6s (u'(s))^2 + s^2 u'(s) -2s- s u(s) -16 u'(s) -4 u(s) u'(s) \right), \nonumber \\
\tilde{C} (s) & = 8\left( 2 +u(s) -s u'(s) -2(u'(s))^2 \right), \nonumber \\
 \tilde{D}(s) & = \frac{2}{3} \Big[ s^2 u''(s) \Big( s^2 u''(s) +2 s u'(s) - 2u(s) \Big)+ s^4 u'(s)^2+4 s^3 u'(s)^3 \nonumber\\
 &-2 s^3 u(s) u'(s)-2 s^3 u'(s) +16 s^2 u'(s)^2+s^2 u(s)^2+2 s^2 u(s) \nonumber\\
 &-10 s u(s)^2 u'(s) -64 s u(s) u'(s)-96 s u'(s)+6 u(s)^3+48 u(s)^2+96 u(s) \Big] \nonumber.
\end{align}
The equation must be solved subject to the $s \to 0^+$ boundary condition 
\begin{align}
\label{e:u1bc} u^{(1)} (s)&= \frac{4} {15} s^2 - \frac{13} {6300} s^4 +\frac{\xi} {1728 \pi} s^5  + {\rm O} (s^6).
\end{align}

\end{prop}

\noindent
{\it Proof.} \quad
We apply the same technique as in Proposition \ref{P1} to the $\tilde{\sigma}$PVI equation \eqref{q5} with parameters \eqref{m4}; that is we change variables $s= \cot X/(2N)$ and we find the terms of order $N^2$ give \eqref{m1a}, while the terms of order $1$ give \eqref{e:u1}. For the boundary condition on $u^{(0)} (s)$, we have \eqref{m2}, which we extend to degree $6$ using \eqref{m1a}. Then we combine this series with \eqref{e:u1} to obtain the boundary condition on $u^{(1)} (s)$. \hfill $\square$

\medskip
One check on the consistency of Proposition \ref{P2} is to verify that substitution of \eqref{e:u1bc} and the first few terms of the power series solution of \eqref{m1a} into \eqref{3.47}, reproduces the expansion (\ref{P13}). We find that this is indeed the case.

\begin{figure}[ht]
 \includegraphics[scale=0.66]{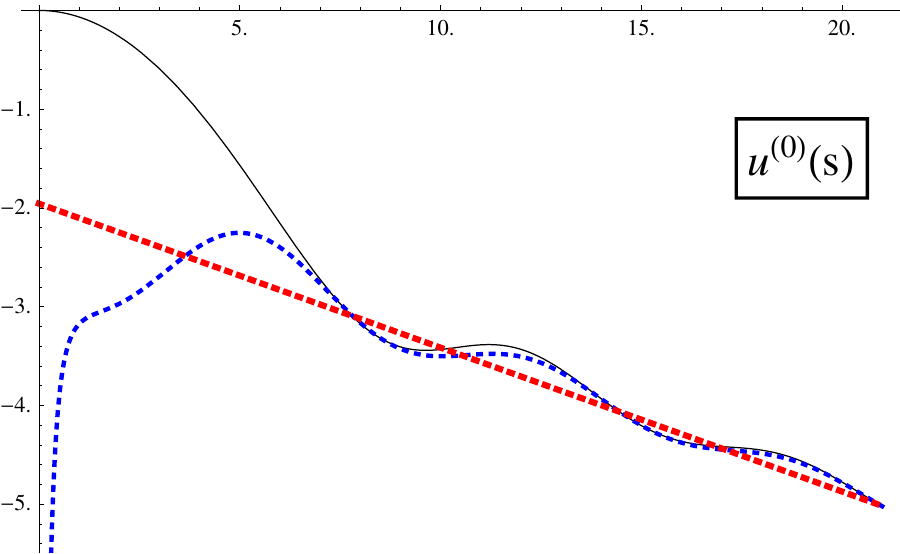} \quad \includegraphics[scale=0.66]{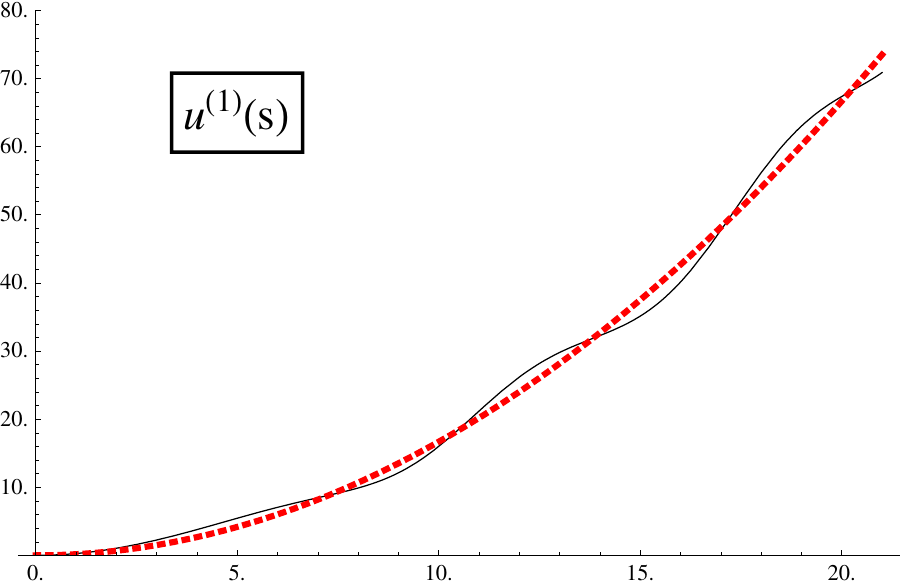}
 \caption{\label{f:uspwfnxi60}$\boldsymbol{\xi=0.6}$. The black curves are numerical solutions of the DEs \eqref{m1a} [left] and \eqref{e:u1} [right] using a sequence of Taylor series of degree $12$. The dashed red curves are the asymptotic approximations \eqref{e:u0larges} [left] and \eqref{e:u1larges} [right]. The dashed blue curve in the graph on the left is a plot of the three leading terms in \cite[eq.~(1.26)]{MT86} with $n=2, \theta=0, k= \frac1{2\pi} \log (1-\xi)$ therein.}
\end{figure}

\begin{figure}[ht]
 \includegraphics[scale=0.66]{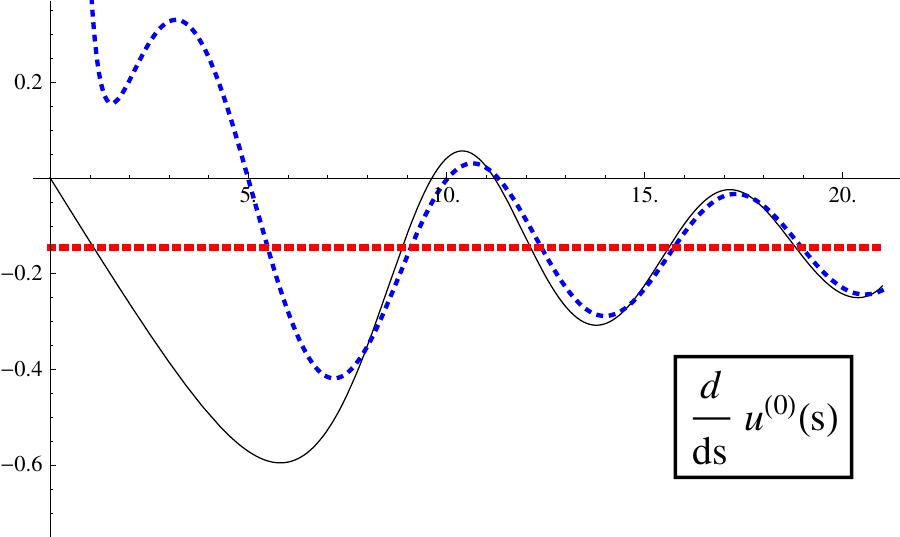} \quad \includegraphics[scale=0.66]{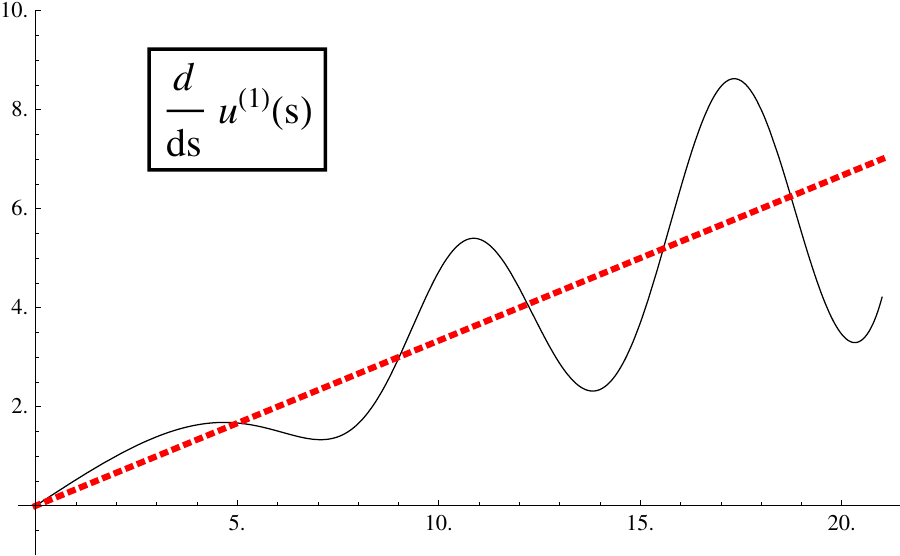}
 \caption{\label{f:udiffs60}$\boldsymbol{\xi=0.6}$. These are the derivatives w.r.t. $s$ of the corresponding curves in Figure \ref{f:uspwfnxi60}.}
\end{figure}

The required consistency between (\ref{3.47}), and (\ref{2.13d}) combined with (\ref{2.13a}), together with knowledge of the large $s$ asymptotic forms in Corollary \ref{Cor3} give information on the large $s$ behaviour of $u^{(0)}(s)$ and $u^{(1)}(s)$. 
In relation to $u^{(0)}(s)$, for $0< \xi < 1$, we have \cite[eq.~(1.26), with $\tilde{\sigma}_2 (s /4) = u^{(0)} (s)$ in our notation]{MT86} 
\begin{align}
\label{e:u0larges} u^{(0)} (s, \xi) = -\frac{k s} {2}+ \frac{k^2} {2} -2 +{\rm O} \left( \frac{1} {s} \right),
\end{align}
(up to a correction $b_0 (n) \to -b_0 (n)$ in \cite[eq.~(1.13)]{MT86}) with $k$ as in \eqref{SS}. From \cite{MT86} we know that the ${\rm O}(1/s)$ term in \eqref{e:u0larges} is oscillatory.
This is clearly displayed in Figure \ref{f:uspwfnxi60}, where we compare a numerical solution of the differential equation \eqref{m1a}, with \cite[eq.~(1.26)]{MT86} 
as given in (\ref{e:u0larges}) with
the oscillatory term O$(1/s)$ included. These oscillatory terms have concrete significance for the deduction of the functional form of $u^{(1)} (s, \xi)$ in the large $s$ limit, as if we are to substitute (\ref{e:u0larges}) for this purpose in 
the differential equation \eqref{e:u1}, we find a result inconsistent with (\ref{SDb}).
Consistency with the latter and (\ref{e:SSsig1}) requires
\begin{align}
\label{e:u1larges} u^{(1)} (s, \xi) = \frac{s^2} {6} + \mathrm{O} (s).
\end{align}
Note that this leading term is independent of $\xi$. From Figure \ref{f:uspwfnxi60} we also see that there are oscillatory sub-leading terms in \eqref{e:u1larges}; the graphs of the derivatives in Figure \ref{f:udiffs60} indicate that these oscillations begin with the O$(s)$ term. 
 
\begin{figure}[ht]
 \includegraphics[scale=0.66]{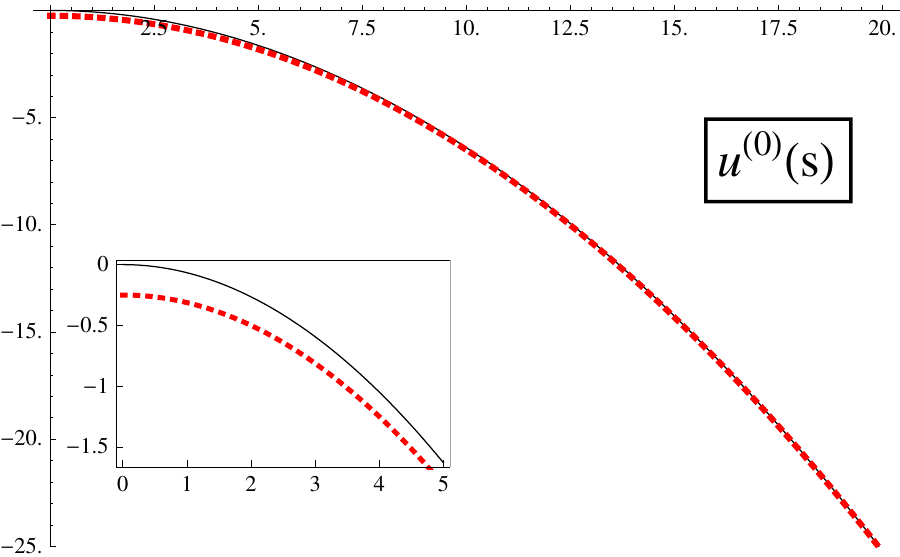} \quad \includegraphics[scale=0.66]{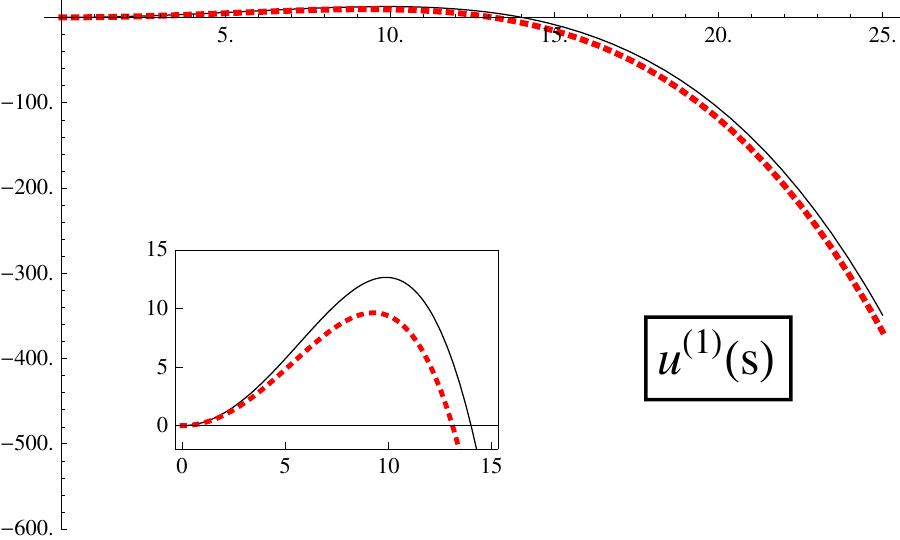}
 \caption{\label{f:uspwfnxi100}$\boldsymbol{\xi=1}$. The black curves are numerical solutions of the DEs \eqref{m1a} [left] and \eqref{e:u1} [right] using a sequence of Taylor series of degree $35$. The dashed red curves are the asymptotic approximations \eqref{e:u0larges100} [left] and \eqref{e:u1larges100} [right]. The insets are the same plots zoomed-in on the origin.}
\end{figure}

For $\xi =1$, we have consistency with (\ref{SDa}) when
\begin{align}
\label{e:u0larges100} u^{(0)} (s, 1) = -\frac{s^2} {16}- \frac{1} {4}+{\rm O} \left( \frac{1} {s} \right),
\end{align}
and substitution of this in (\ref{e:u1}) gives
\begin{align}
\label{e:u1larges100} u^{(1)} (s, 1) = -\frac{s^4} {768}+ \frac{43 s^2} {192}+{\rm O} \left( 1 \right);
\end{align}
see Figure \ref{f:uspwfnxi100} for comparisons of these asymptotic forms against numerically generated solutions
of the corresponding differential equations.

\section{Numerical power series solution of the differential equations and evaluation of the spacing distributions}
\label{s:Num}

The utility of Propositions \ref{P1} and \ref{P2} is that we can use a power series method together with computer packages to numerically compute to high accuracy the next-to-leading order term in the nearest neighbour spacing distribution (\ref{e:ps0xi}); this can be done directly by computing the terms in (\ref{3.47}) or via the second derivative of (\ref{C1}). Such power series methods were introduced as a technique to compute Painlev\'e transcendents associated with spacing distributions in random matrix theory in \cite{PS03}. This was in relation to the soft edge. Subsequently the same numerical method was used for the computation of bulk spacings \cite[\S 8.3.4]{Fo10}. The first point is to note that in generating a power series solution of the $\sigma$PV equation (\ref{d1}) about the origin, one finds a radius of convergence of approximately $8.5$ (for both $\xi=1$ and $\xi =0.6$). Inside the radius of convergence, this power series can be used to accurately evaluate the transcendent and its derivative, and from this data a new power series can be computed and the procedure iterated to cover the interval from $s=0$ to beyond $s=20$. With $\sigma^{(0)}(s)$ so determined,
iterative power series solutions of the differential equation in Proposition \ref{P1} can be obtained.
It is through this procedure, and its analogue starting with the differential equation (\ref{m1a}) and proceeding to the
differential equation of  Proposition \ref{P2}, that the graphs of $\sigma^{(0)}(s), \sigma^{(1)} (s), u^{(0)} (s), u^{(1)} (s)$
displayed in Figures \ref{f:pwfnxi100}--\ref{f:uspwfnxi100} have been generated.

We remark that use of a power series method to generate solutions over a large interval seems necessary
in the cases $0< \xi < 1$. In particular, if in these cases one tries to use a computer algebra package to
solve the differential equation (\ref{d1}) subject to 
initial conditions for $\sigma^{(0)} (x_0)$ and $\sigma^{(0)} {}' (x_0)$ with $x_0$ small,  as computed from the boundary
condition (\ref{d2}) (with the latter further extended in accuracy as given in (\ref{Sd0})), it is found that as $x$ is increased
from $x_0$, the DE solver gives incorrect values, and furthermore soon diverges to a spurious pole.

\begin{figure}[ht]
 \includegraphics[scale=0.45]{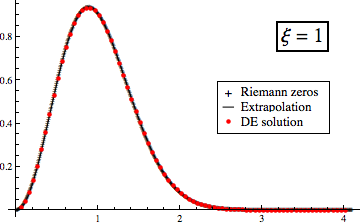} \quad \includegraphics[scale=0.45]{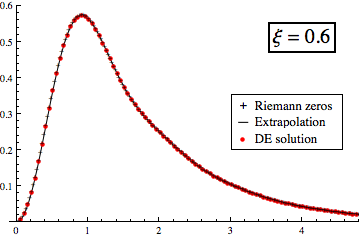} 
 \caption{\label{f:p0p1}Comparison of Riemann zero nearest neighbour spacing (where each zero is scaled by 
 the leading term in \eqref{1A}) with the corresponding DE solution (\textit{ie}. the second derivative of \eqref{2.13d}). We also compare these to an extrapolated limit of a sequence of finite calculations as in \eqref{2.13a}.}
\end{figure}

\begin{figure}[ht]
 \includegraphics[scale=0.45]{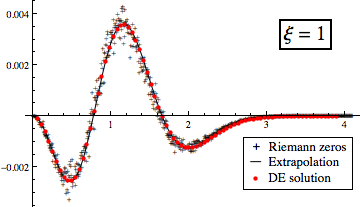} \quad \includegraphics[scale=0.45]{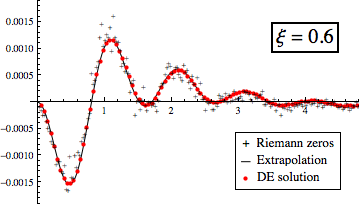} 
 \caption{\label{f:p1}Comparison of next-to-leading order corrections to the quantities in Figure \ref{f:p0p1}.}
\end{figure}

Substitution of the piecewise  functions for $\sigma^{(0)} (s)$ and $\sigma^{(1)} (s)$ into (\ref{2.13d}) gives us an approximation of $\det ( \mathbb I - \xi \mathbb K_s^N)$ up to the first order correction, and substitution into (\ref{C1}) isolates the correction term. Taking the second derivative with respect to $s$ we obtain Figures \ref{f:p0p1} and \ref{f:p1} respectively. Note that to obtain the comparison with the Riemann zero data in these figures, we have used the scalings (\ref{e:Neff}) and (\ref{e:alphas}) from \cite{BBLM06}. Also in Figures \ref{f:p0p1} and \ref{f:p1} we plot an extrapolation of a sequence of finite calculations \eqref{2.13a}, similar to that in \cite{BBLM06}: by calculating \eqref{2.13a}, with kernel \eqref{2.0}, for 20 values of $N$ between 100 and 138 we extrapolate the limiting value of the nearest neighbour spacing and the next-to-leading order correction at each point. 
To graphical accuracy the extrapolated values are identical to those computed from the differential equation.
Note that when $\xi=0.6$ the correction term shows more pronounced oscillations for finite $s$ values than its $\xi = 1$
counterpart, a feature that seems to be be driven by oscillations in the functional forms for the corresponding Painlev\'e
transcendents.

Most strikingly, the predicted functional forms from random matrix theory show accurate agreement with the 
Riemann zero data both for $\xi = 1$ \cite{BBLM06}, and upon thinning of the Riemann zero data set with $\xi = 0.6$,
for all displayed values of $s$. This is in contrast to what was found in Figure \ref{f:2ptcomp} for the two-point function,
where the accuracy deteriorates after approximately one and a half units of the mean spacing. A significant difference between the two quantities that may explain this observation relates to the respective large $s$ forms. That is, the spacing distribution decays as an exponential or faster (see \eqref{3.47} with \eqref{e:u0larges}--\eqref{e:u1larges100}), whereas the correction term for the two-point correlation function is an oscillatory order one quantity for all
$s$ (see \eqref{Sc}). It follows that at a graphical level discrepancies in the case of the correction term to the
spacing distribution functional formal form cannot be quantitatively probed in distinction to the situation with the correction term
for the two-point correlation function. 

In conclusion, our study corroborates the study of \cite{BBLM06}, and so lends further weight to the conjecture
that correction terms to the Montgomery--Odlyzko law themselves have a random matrix interpretation. Specifically,
as put forward in \cite{BBLM06} as an extension of the earlier work \cite{KS00a}, the results of our study are
consistent with the hypothesis that the correction terms to the Montgomery--Odlyzko law for correlation functions and
associated distribution functions of the Riemann zeros coincides with the  ${\rm O}(1/N^2)$ correction terms for the 
corresponding quantities of the eigenvalues of random unitary matrices. This holds with the relationship between $E$ and
$N$ given by \eqref{e:Neff}, and with the change of scale (\ref{e:alphas}).

\section*{Acknowledgements}
We are very appreciative of A.~Odlyko in providing us with the Riemann zeros data set used in our work. We also thank F.~Bornemann for spotting some points for correction in an earlier draft. This research project is part of the program of study supported by the ARC Centre of Excellence for Mathematical \& Statistical Frontiers.

\providecommand{\bysame}{\leavevmode\hbox to3em{\hrulefill}\thinspace}
\providecommand{\MR}{\relax\ifhmode\unskip\space\fi MR }
\providecommand{\MRhref}[2]{%
  \href{http://www.ams.org/mathscinet-getitem?mr=#1}{#2}
}
\providecommand{\href}[2]{#2}

\end{document}